\documentclass[twocolumn,apj]{emulateapj}
\usepackage{graphics}
\usepackage{epstopdf}
\usepackage{epsfig}
\usepackage{amsmath}
\usepackage{amssymb}

\shortauthors{Park et al.}
\shorttitle{Particle-in-cell simulations of particle energization via shock drift acceleration}

\begin{document}

\title{Particle-in-cell simulations of particle energization via shock drift acceleration from low Mach number quasi-perpendicular shocks in solar flares} 

\author{Jaehong Park\altaffilmark{1,2}, Chuang Ren\altaffilmark{1,2,3}, Jared C. Workman\altaffilmark{4,1}, and Eric G. Blackman\altaffilmark{1,2}}
\affil{$^1$Department of Physics \& Astronomy, University of Rochester, Rochester NY, 14627}
\affil{$^2$Laboratory for Laser Energetics, University of Rochester, Rochester NY, 14623 }
\affil{$^3$Department of Mechanical Engineering, University of Rochester, Rochester NY, 14627}
\affil{$^4$Department of Physical \& Environmental Sciences, Colorado Mesa University, Grand Junction CO, 81501}

\begin{abstract}
Low Mach number, high beta fast mode shocks can occur in the magnetic reconnection outflows of solar flares. These shocks, which occur above flare loop tops, may provide the electron energization responsible for some of the observed hard X-rays and contemporaneous radio emission.  
Here we present new 2D particle-in-cell simulations of low Mach number/high beta quasi-perpendicular shocks. The simulations show that electrons above a certain energy threshold experience shock-drift-acceleration. The transition energy between the thermal and non-thermal spectrum and the spectral index from the simulations are consistent with some of the X-ray spectra from RHESSI in the energy regime of $E\lesssim 40\sim 100$ keV. 
Plasma instabilities associated with the shock structure such as the modified-two-stream and the electron whistler instabilities are identified using numerical solutions of the kinetic dispersion relations. We also show that the results from PIC simulations with reduced ion/electron mass ratio can be scaled to those with the realistic mass ratio.
\end{abstract}


\section{Introduction}

Low Mach number ($M$), high plasma beta ($\beta_p$) fast mode shocks can occur in the magnetic reconnection outflows of solar flares. Hard X-ray data from Yohkoh and RHESSI has revealed that electrons are energized above flare loop tops and foot points \citep[e.g.,][]{lin03}.
Solar flares are diverse and the associated reconnection events are likely 
``acceleration environments" with potentially different mechanisms of particle acceleration operating on different scales.  One potential source of particle acceleration, seen from  analytic predictions \citep{blackman94} and numerical simulations  of reconnection configurations in which an obstacle was present \citep[e.g.,][]{forbes88,workman11}, involves the presence of low Mach number fast shocks in  reconnection outflows.  These ``termination"  shocks may contribute to the 
high energy acceleration observed over some frequency range,  studying the potential ways in which these shocks can accelerate particles is well motivated.

\citet{mann06,mann09} and \citet{warmuth09} suggested an electron energization via shock drift acceleration (hereafter SDA) in termination shocks.
\citet{guo10,guo12} performed hybrid simulations for termination shocks where test electrons were effectively energized via the interaction with pre-existing large-scale magnetic fluctuations.
In our recent work \citep{park12}, we performed a full particle-in-cell (PIC) simulation for purely perpendicular (i.e. magnetic field perpendicular to the shock normal), low \textit{M}/high $\beta_p$ shocks. We found that both electrons and ions participated in SDA. 

Termination shocks may however deviate from purely perpendicularity in solar flares, the extent to which is an open  question. In the meantime, it is instructive to relax the constraint that the shocks are exactly perpendicular and consider the case of quasi-perpendicularity. The difference is significant because particles can cross back upstream for quasi-perpendicular shocks.
There were analytical \citep{wu84,krauss89b} and hybrid simulation studies \citep{krauss89a} of SDA in a nearly perpendicular bow shock for energetic electrons.
Recently, \citet{matsukiyo11} performed 1D PIC simulation for the electron SDA in low Mach number quasi-perpendicular shocks in galaxy clusters.

In this paper, we present the results of 2D PIC simulations for quasi-perpendicular low $M$/high $\beta_p$ shocks in solar flares. The upstream magnetic field makes  angles of $\theta_B=80^\circ,82^\circ$, and $83.5^\circ$ respectively to the shock normal and the shocks satisfy the subluminal condition, $V_{sh}/\text{cos}\theta_B<c$, where $V_{sh}$ is the shock speed in the upstream rest frame and $c$ is the speed of light.
In such a subluminal shock, electrons can then be reflected at the shock front due to the magnetic mirror effect and gain energy \citep[e.g.,][]{ball01,mann06,mann09,warmuth09}.

One difficulty in performing PIC simulations for SDA in solar flares
is that they require sufficiently high energy electrons
above the threshold energy for SDA in the simulation to obtain statistically reliable results. The commonly used Maxwellian distribution has  too few above the threshold electrons to be used in a simulation. Herein we use a kappa distribution with $\kappa=10$ for both injected ions and electrons. This increases the number of high energy electrons but does not significantly alter the original shock structure, which is determined by the bulk of the thermal particles. 
Although the kappa distribution is used for computational convenience in this paper, the kappa distribution may be physically a relevant distribution in the outflow driven electron-ion jet \citep[e.g.,][]{yoon06,kasparova09,mann06,mann09,warmuth09}.

The goals of this paper are two-fold: 
(1) studying the formation and structure of such shocks, including the turbulent dissipation mechanism 
for collisionless shock sustenance and entropy creation,
and (2) studying the particle acceleration mechanism relevant for the soft/hard X-ray flux observations in solar flares. 
We observe the same modified-two-stream instability \citep{krall71} in the shock transition region as in the perpendicular shocks \citep{park12} that can provide the turbulent dissipation in the downstream. Furthermore, a temperature anisotropy after the shock transition region is found to drive the electron whistler instability \citep[e.g.,][]{gary93}.

As shown below, more SDA-accelerated electrons are found in the present simulations of the quasi-perpendicular case compared to the previously simulations of the perpendicular case \citep{park12}.
In this context, we also  extend the theoretical analysis in \citet{mann06,mann09,warmuth09} to include the electric potential jump at the shock front \citep[e.g.,][]{ball01} and generalize the electron energy spectrum.
Both our theoretical analysis and our simulations show a transition energy, $E_\text{trans,p}$, between the thermal and non-thermal photon spectrum that is determined by the minimum angle of $\theta_B$. 
We find that the transition energy $E_\text{trans,p}$ and the spectral index $\delta$ from theory and our simulation are consistent with some of the X-ray spectra of solar flares from RHESSI \citep[e.g.,][]{altyntsev12} in the energy range of $E\lesssim 40\sim 100$keV.
Beyond this range, to maintain the power-law distribution up to $E\sim$MeV, additional mechanisms beyond the shock acceleration studied are required.

The rest of the paper is organized as follows. The simulation setup is described in Section \ref{setup}. The shock structure and particle acceleration are described in Section \ref{result}. We summarize in Section  \ref{summary}.

\section{Simulation Setup}
\label{setup}

We use the fully-relativistic full PIC code OSIRIS \citep{fonseca02} to study the formation of and particle energization in low-$M$,  high-$\beta_p$ shocks, where $M$ is the Mach number and $\beta_p$ is the
ratio of thermal to magnetic pressure.  To launch a shock, we use the moving wall boundary condition \citep{langdon88,park12} at the right boundary of the 2D simulation box.  The moving wall method generates a slowly propagating shock compared to the more standard fixed reflection boundary method, and allows for smaller box sizes and more efficient use of simulation time.

We adopt  parameters typical of those found in solar flare reconnection outflows  \citep{tsuneta96,workman11} as the upstream conditions for our shock. In particular, we use a  plasma density $n=5\times 10^9/\text{cm}^3$, electron and ion temperatures $T_e=T_i=0.8\text{keV}(=9.27\times 10^6\text{K})$, and the magnetic field strength $B=6$G with $\beta_p\equiv 8\pi n(T_e+T_i)/B^2=8.93$. 
The magnetic field is in the $x$-$y$ plane $({\bf B}=B_x\hat{x}+B_y\hat{y})$ and has an angle of $\theta_B=80^\circ$, $82^\circ$, and $83.5^\circ$ from the shock normal($x$-axis) in each simulation.
We also performed another simulation where the upstream magnetic field 
is directed out of the simulation plane $(B=B_x\hat{x}+B_z\hat{z})$ and has an angle of $\theta_B=80^\circ$ to compare the electron energy spectrum with that in the ``in-plane" $B$-field simulation.
A reduced ion/electron mass ratio of $m_i/m_e=30$ is used to reduce  computational demands. The Alfv\'en Mach number is chosen to be $M_A\equiv V_1\sqrt{4\pi m_in}/B=6.62$, which equates to an upstream plasma flow velocity in the shock rest frame, $V_1=0.032c$, for $m_i/m_e=30$. If the real mass ratio, $m_i/m_e=1836$, is used for the same $M_A$, then the upstream flow velocity would be $1226$km/s$(=0.0041c)$.  Effects of a realistic mass ratio are further discussed in Section 3.3. 
The super-fast-magnetosonic Mach number $M$ satisfies  $M\equiv M_A/\sqrt{1+(5/6)\beta_p}=2.28$. The ratio of the electron cyclotron frequency to the electron plasma frequency is $\Omega_{ce}/\omega_{pe}=0.0265$.
With these upstream values of $M_A$ and $\beta_p$, the Rankine-Hugoniot relation \citep{tidman71} for the shocks with the angles used gives a compression ratio in the range of  (2.06, 2.50), where the lower and the upper limits are calculated for 2D and 3D, respectively.

\begin{figure}
\begin{center}
\includegraphics[scale=.45]{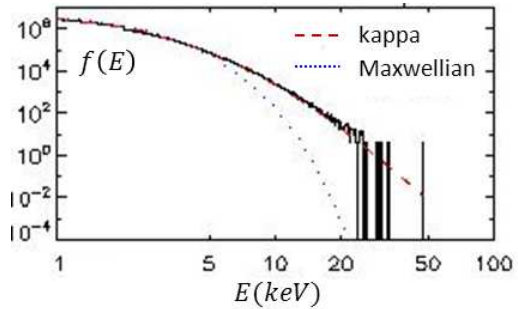}
\caption{The initial energy distribution in the upstream rest frame from the simulation. The dashed line is a kappa distribution with $\kappa=10$ and $T=0.8$keV. The dotted line is a Maxwellian distribution with $T=0.8$keV.}
\label{kappadist}
\end{center}
\end{figure}

The simulation box is initialized with a kappa--distributed ion-electron plasma drifting with $V_d=0.0213c$ and $T_e=T_i=0.8\text{keV}$,
where $V_d$ is set to a smaller value than the upstream speed $V_1(=0.032c)$ in the shock rest frame anticipating that the shock will be traveling to the left. In the upstream rest frame, the energy distribution is a kappa distribution such as
\begin{eqnarray}
f(E)={2^{5/2}\over\sqrt{\pi}}{1\over(2\kappa-3)^{3/2}}
{\Gamma(\kappa+1)\over\Gamma(\kappa-1/2)}
{E^{1/2}\over T^{3/2}} \nonumber \\
\times\left(1+{2\over 2\kappa-3}{E\over T}\right)^{-\kappa-1},
\label{kappa}
\end{eqnarray}
where $\Gamma$ is the gamma function and  
$f(E)$ is normalized to $1$, $\int_0^{\infty}dE f(E)=1$. The kappa distribution approaches the Maxwellian distribution 
as $\kappa$ goes to $\infty$.
Figure \ref{kappadist} shows the initial energy distribution in the upstream rest frame from the simulation. The dashed line is a kappa distribution in Equation (\ref{kappa}) with $\kappa=10$ and $T=0.8$keV and the dotted line is a Maxwellian distribution with the same temperature. The bulk parts ($E<5$ keV) of the two distributions are very similar. The implementation of the kappa distribution in OSIRIS is described in Appendix \ref{app1}.

\begin{figure*}
\begin{center}
\includegraphics[scale=.38]{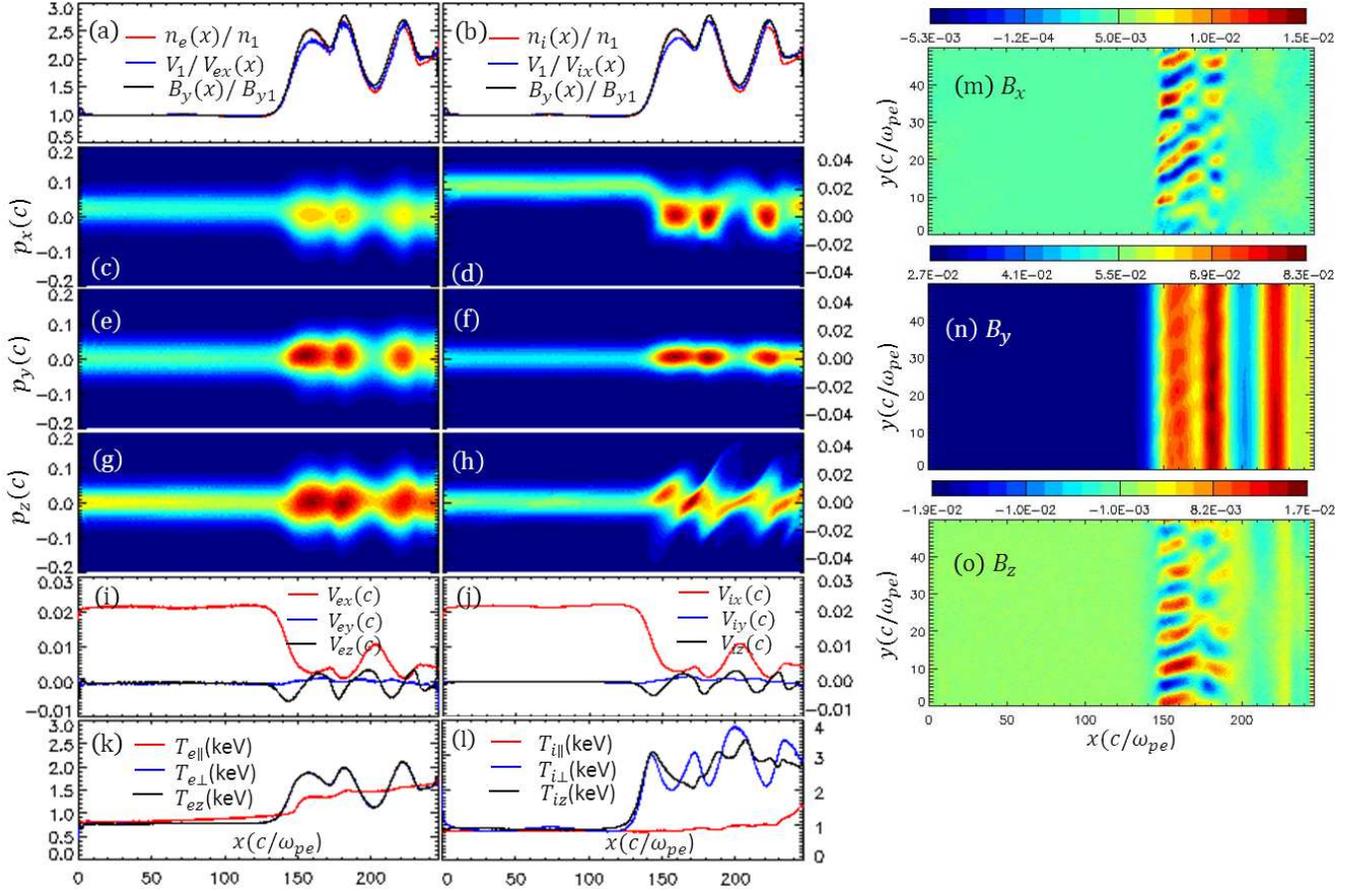}
\caption{The ratios, $n/n_1$, $V_1/V_x$, and $B_y/B_{y1}$,
momentum distribution of $p_x-x$, $p_y-x$, and $p_z-x$, $y$-averaged flow velocity of $V_x$, $V_y$, and $V_z$, and temperature $T_\parallel$, $T_\perp$, $T_z$ for the electrons(left column) and the ions(right column) at $t=11200/\omega_{pe}$.
$(a)\sim(b)$ are calculated in the shock-rest frame and the other plots are obtained in the simulation frame. (m-o) show the $B_x$, $B_y$, and $B_z$ fields in the $x-y$ space.}
\label{phase}
\end{center}
\end{figure*}

A uniform external $E_z$ field is set up along the $z$-axis with $E_z=-V_dB_y/c$. A new plasma of the same kappa distribution is constantly injected from the left boundary ($x=0$) throughout the simulation. The simulation box sizes are $L_x\times L_y=250c/\omega_{pe} \times 50 c/\omega_{pe}$.
The grid size used is $dx=dy=0.08c/\omega_{pe}$ and the time step used is $dt=0.056/\omega_{pe}$. For each particle species, $196$ particles per cell are used. The particle number/cell is fairly large to reduce numerical collisions and maintain the kappa distribution for a sufficiently long time.

A linear current deposition scheme is used for all simulations in this paper.
A periodic boundary condition is used in the $y$-direction for both particles and fields. For fields, an open boundary condition is used in the $x$-direction. Particles that reach  $x=0$ are re-injected into the box with the initial kappa distribution. At $x=L_x$, a moving wall boundary condition is adopted \citep{park12} and the moving wall speed is set to $V_\text{wall}=0.004c$.

\section{Results and analysis}
\label{result}
\subsection{Shock structure}

In this subsection, we present results for the shock structure for $\theta_B=80^\circ$ in-plane case only
but other simulations for $\theta_B=82^\circ$ and $83.5^\circ$ also show similar shock structures.  The electron energization via SDA shows different energy spectra for the different angles of $\theta_B$ as will be seen in the next subsection.

In Figure \ref{phase}(a\&b), we plot for both species the upstream to downstream ratios of
density  $n(x)/n_1$,   flow speeds $V_1/V(x)$, and  $y$-magnetic field, $B_y(x)/B_{y1}$.
In Figure  \ref{phase}(c-l), we plot the phase-space distributions of 
$p_x-x$, $p_y-x$, and $p_z-x$, the $y$-averaged flow velocity profiles of
$V_x(x)$, $V_y(x)$, and $V_z(x)$, and the temperature profiles of
$T_\parallel(x)$, $T_\perp(x)$, and $T_z(x)$
(where $\perp$ and $\parallel$ are perpendicular and parallel to the magnetic field $\bold{B}(x)\equiv B_x(x)\hat{x}+B_y(x)\hat{y}$, respectively), for the electrons (the left column) and the ions (the right column) at $t=11200/\omega_{pe}$ for the $\theta_B=80^\circ$ case.

In Figure \ref{phase}, the shock front is at $x\approx 140c/\omega_{pe}$ and
moves to the left with a speed of $0.01c$ in the simulation frame.
In Figure  \ref{phase}(a\&b), the compression ratio is $r=2.5$ near the shock front and oscillates around $2.0$  downstream, in reasonable agreement with the Rankine-Hugoniot relation \citep{tidman71}.
Figure \ref{phase}(c-h) show that the electrons and ions are heated  downstream but the ions are conspicuously less heated in the $y$-direction [Figure \ref{phase}f].
In Figure \ref{phase}(i\&j), the flow velocity, $V_x=0.021c$, in the upstream
decreases downstream to the moving wall speed, $V_\text{wall}=0.004c$.  
The $z$-component of the velocity, $V_z$, is oscillating around zero downstream due to the $\bold{E}_x\times\bold{B}_y$-drift while $V_y\simeq  0$. 
Figure \ref{phase}(k\&l) show that the downstream temperature is  anisotropic, that is $T_\perp>T_\parallel$ 
with $T_{e\perp}=T_{ez}\approx 1.66$keV and $T_{e\parallel}\approx 1.5$keV for the electrons, and $T_{i\perp}=T_{iz}\approx 2.5$keV and $T_{i\parallel}\approx 1.0$keV for the ions.

\begin{figure*}
\includegraphics[scale=.34]{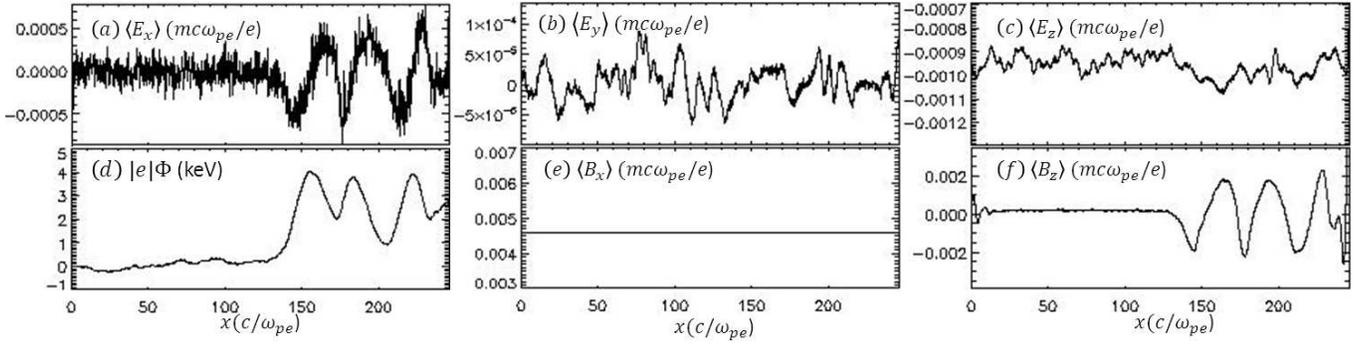}
\caption{$y$-averaged $\bold{E}$ and $\bold{B}$ fields, and the potential energy $|e|\phi(x)$ at $t=11200/\omega_{pe}$ measured in the shock rest frame.}
\label{potenital}
\end{figure*}

\begin{figure*}
\begin{center}
\includegraphics[scale=.36]{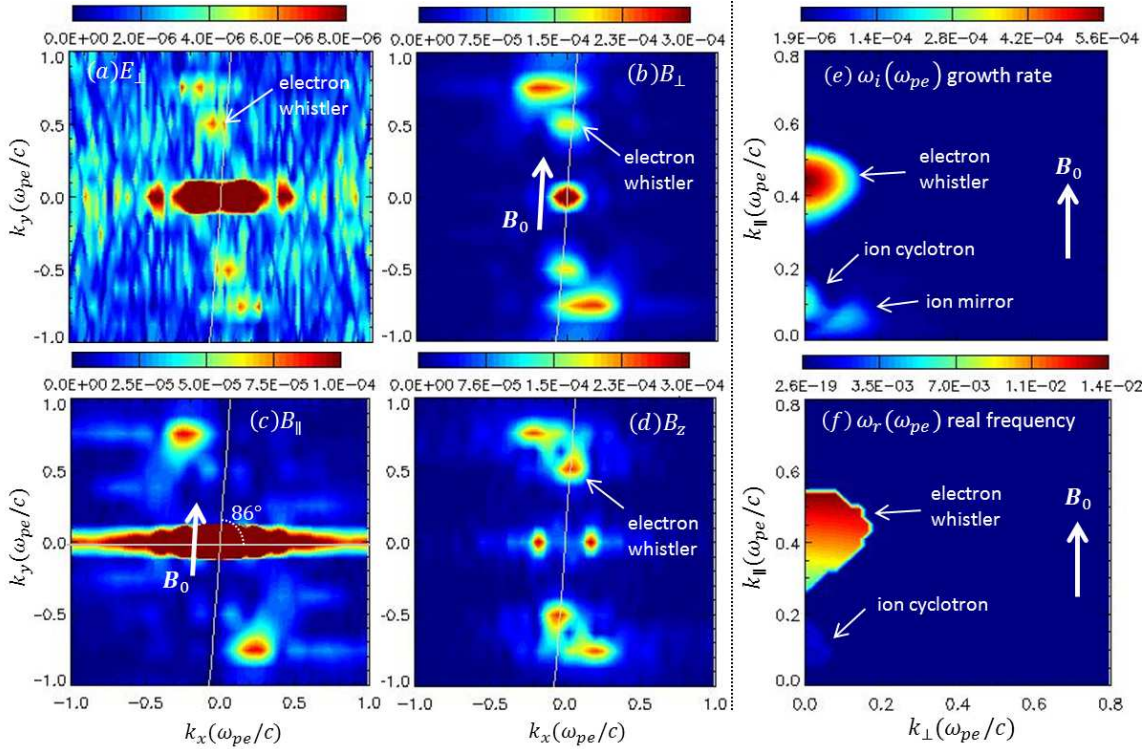}
\caption{(a)$\sim$(d): Fourier spectra of the $\bold{E}$ and $\bold{B}$ fields in the $k_x-k_y$ space from the simulation.
(e): Numerically solved growth rates of temperature anisotropy-driven instabilities in the region, $140<x<170(c/\omega_{pe})$ of Figure \ref{phase}.}
\label{fft}
\end{center}
\end{figure*}

Figure \ref{phase}(m-o) show  $\bold{B}$  in the 
$x-y$ space. The $B_x$ and $B_z$ fields show oscillatory patterns along the $y$-axis near the shock front ($x\sim 150c/\omega_{pe}$) and the $B_y$ field shows a rippled surface along the $y$-axis. These variations are due to electron temperature anisotropy-driven instabilities, $T_{e\perp}>T_{e\parallel}$.

Figure \ref{potenital} shows the $y$-averaged $\bold{E}$, $\bold{B}$, 
and the potential energy $|e|\Phi(x)(=e\int_{x_1}^{x}E_xdx)$ at $t=11200/\omega_{pe}$, measured in the shock rest frame. The $E_y$, $E_z$, and $B_x$ fields are approximately constant across the shock while $E_x$ and $B_z$ oscillate  with a wave number $k=0.2\omega_{pe}/c$  downstream.
The potential energy jump $|e|\Delta\Phi$ at the shock front is $\sim 3.5$keV [Figure \ref{potenital}d].
Some ions in the low energy tail reflect at the shock front and can drive the modified two-stream instability. 

In Figure \ref{fft}(a-d), we plot the Fourier spectra of the $E_\perp$, 
$B_\perp$, $B_\parallel$, and $B_z$ fields, 
where $\perp$ and $\parallel$ are perpendicular and parallel to $\bold{B}_0$, respectively. The $\bold{B}_0(=B_{0x}\hat{x}+B_{0y}\hat{y})$ is an averaged field over $140<x< 170(c/\omega_{pe})$ in Figure \ref{phase} and has an angle $86^\circ$ from the $x$-axis.
Two distinct dominant modes are excited by temperature anisotropy-driven instabilities $(T_\perp>T_\parallel)$, one is along the $\bold{B}_0$ axis with 
$k=0.5\omega_{pe}/c$ and another is sightly deviated from the $\bold{B}_0$ axis with $k=0.8\omega_{pe}/c$. 
In Figure \ref{fft}(d), the strong signals at $k_x=\pm 0.2\omega_{pe}/c$ are due to the oscillatory pattern of the $B_z$ field along the $x$-axis as seen in Figure \ref{potenital}(f).

To analyze the spectra from the simulation, 
we numerically solve the dispersion relation \citep[e.g.,][]{gary93} in Figure \ref{fft}(e\&f). Here we assume bi-Maxwellian electron and ion distributions and a uniform background magnetic field $B_0$. We use the parameters extracted from the simulation with $B_0=13.5G$, $T_{e\parallel}=1.25$keV and $T_{e\perp}=1.66$keV for the electrons, and $T_{i\parallel}=0.86$keV and $T_{i\perp}=2.55$keV for the ions, where $\parallel$ and $\perp$ are parallel and perpendicular directions to $B_0$, respectively. 

In Figure \ref{fft}(e), the electron whistler modes are centered on the $B_0$ axis with $k=0.45\omega_{pe}/c$, the ion cyclotron modes centered on the $B_0$ axis with $k=0.1\omega_{pe}/c$, and the ion mirror modes are obliquely off the $B_0$ axis with $k=0.12\omega_{pe}/c$.
The maximum growth rates are $\omega_{i}=5.6\times 10^{-4}\omega_{pe}$, $2.5\times 10^{-4}\omega_{pe}$, and $2.0\times 10^{-4}\omega_{pe}$ for the electron whistler, the ion cyclotron, and the ion mirror modes, respectively. In Figure \ref{fft}(f), the real frequencies for the electron whistler and the ion cyclotron modes are around $\omega_{r}=0.012\omega_{pe}$ and $0.0012\omega_{pe}$, respectively, 
and the ion mirror modes have zero real frequency.
These are consistent with the analytical results \citep[e.g.,][]{gary93},
$\Omega_{ci}<\omega_r<|\Omega_{ce}|$ for the electron whistler modes and $0<\omega_r<\Omega_{ci}$ for the ion cyclotron modes.

The linear theory based on bi-Maxwellian distributions with the parameters found in the PIC simulations show that electron mirror modes are marginally stable \citep[e.g.,][]{gary06}. The oblique modes found in Figure \ref{fft}(a$\sim$d)] could be due to some unknown modes. However, given the fact that the actual distribution is not exactly a bi-Maxwellian (e.g., with flows) and the uncertainty in the temperature anisotropy measurements in the simulation, the oblique modes with $k=0.8\omega_{pe}/c$ seen in the simulation [Figure \ref{fft}(a$\sim$d)] may possibly still be a electron mirror-type of  modes for a non-bi-Maxwellian distribution. In addition, the ion cyclotron/mirror modes are not observed in this simulation because of the small box size $L_y=50c/\omega_{pe}$ and/or their relatively weaker growth.

\begin{figure}
\begin{center}
\includegraphics[scale=.35]{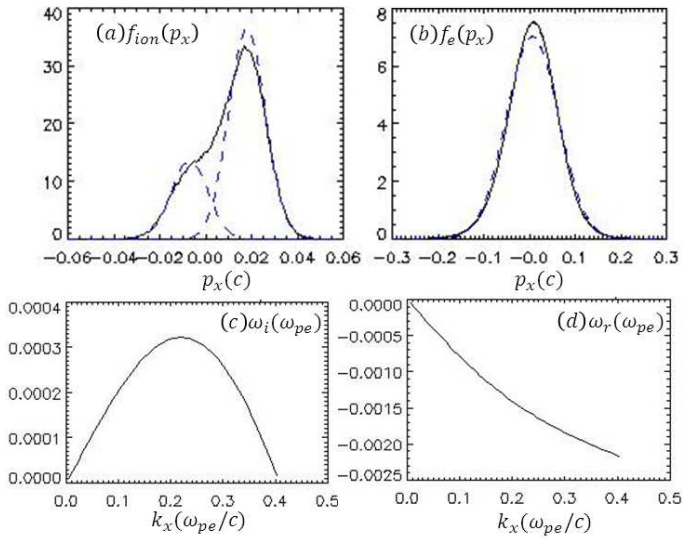}
\caption{(a\&b): Ion and electron distributions in the shock transition region, $135<x<155(c/\omega_{pe})$ at $t=11200/\omega_{pe}$. We fit the distributions with  Maxwellian distributions (dashed lines).
(c\&d): The growth rates and real frequencies obtained from the MTSI dispersion relation (Appendix \ref{app2}).}
\label{fdist}
\end{center}
\end{figure}

Figure \ref{fdist}(a\&b) shows the electron and ion distributions in the shock transition region in $135<x<155(c/\omega_{pe})$ at $t=11200/\omega_{pe}$ from the simulation. 
We observe that $27\%$ of the incoming ions are reflected at the shock front [Figure \ref{fdist}a] and the modified two-stream instability (MTSI)
\citep[e.g.,][]{matsukiyo03,matsukiyo06,umeda10,umeda12} can be excited. 
Here we solve the MTSI in the electrostatic limit with its wave vector along the $x$-axis (Appendix \ref{app2}).
The electrons have a temperature of $T_{ex}=1.65$keV and drift with $V_{ex}=0.0064c$ in the simulation frame [Figure \ref{fdist}b].
We fit the distributions using drifting Maxwellians.
The electrons are magnetized with $B=10G$ while the ions are assumed to be non-magnetized. In the electron rest frame, the drift velocities for the incoming and reflecting ions are $V_{x\text{in}}=0.0116c$ and $V_{x\text{re}}=-0.0134c$, respectively. Both incoming and reflecting ions have a temperature of $T_{x\text{in}}= T_{x\text{re}}=0.98$keV.
In Figure \ref{fdist}(c\&d), we solve the kinetic dispersion relation for the MTSI. The maximum growth rate is $\omega_i=3.2\times 10^{-4}\omega_{pe}$ at $k=0.2c/\omega_{pe}$ and the real frequency is $\omega_r=-0.0015\omega_{pe}$.

Turbulent dissipation is needed to randomize the upstream flow to form collisionless shocks \citep[e.g.,][]{wu82}. 
The macroscopic jump conditions across a shock are essentially independent of the source of microphysical turbulence, as long as there is such a source.
Plasma instabilities in the shock transition region are natural  sources for this needed turbulent dissipation and we consider  three possible instabilities: 
(1)lower hybrid instability (the excited modes are in the $\bold{B}\times\nabla\bold{B}$ direction) \citep[e.g.,][]{zhou83}, (2)whistler instability, and (3)MTSI \citep[e.g.,][]{papadopoulos71,wagner71}.
The lower hybrid instability is precluded in these 2D in-plane simulations where the $\bold{B}$-field is in the $x-y$ simulation plane. 
In addition, we notice that the shock structure (i.e. the compression ratio) in the simulation with an in-plane $B$ field is the same as that with the out-of-plane $B$ field which then precludes the whistler instability. 
We are led to conclude that  the MTSI is the most likely candidate to provide the needed turbulence \citep[e.g.,][]{park12}.

\subsection{particle heating via shock drift acceleration}

\begin{figure*}
\begin{center}
\includegraphics[scale=.37]{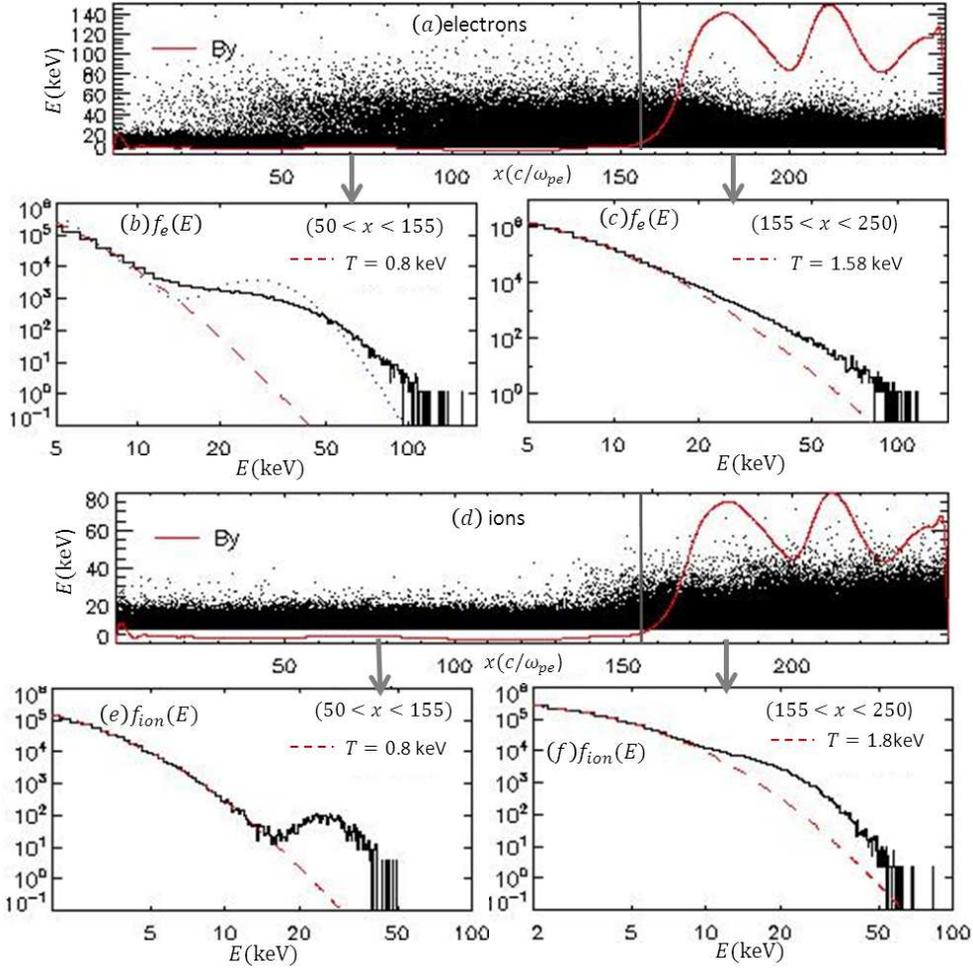}
\caption{(a\&d): Energy distribution vs. $x$-ranges where the $B_y$ field is over-plotted with an arbitrary scale.
The electron(b\&c) and ion(e\&f) energy distributions in upstream 
 $(50<x<155 c/\omega_{pe})$ and  downstream $(155<x<250 c/\omega_{pe})$ at $t=8400/\omega_{pe}$. 
In (b),
 the dotted line is a theoretical energy distribution when the potential energy at the shock front is $e\Phi=-3.5$(keV).
We fit the thermal distributions with kappa distributions with $\kappa=10$ (dashed lines).}
\label{edist}
\end{center}
\end{figure*}

\begin{figure*}
\begin{center}
\includegraphics[scale=.35]{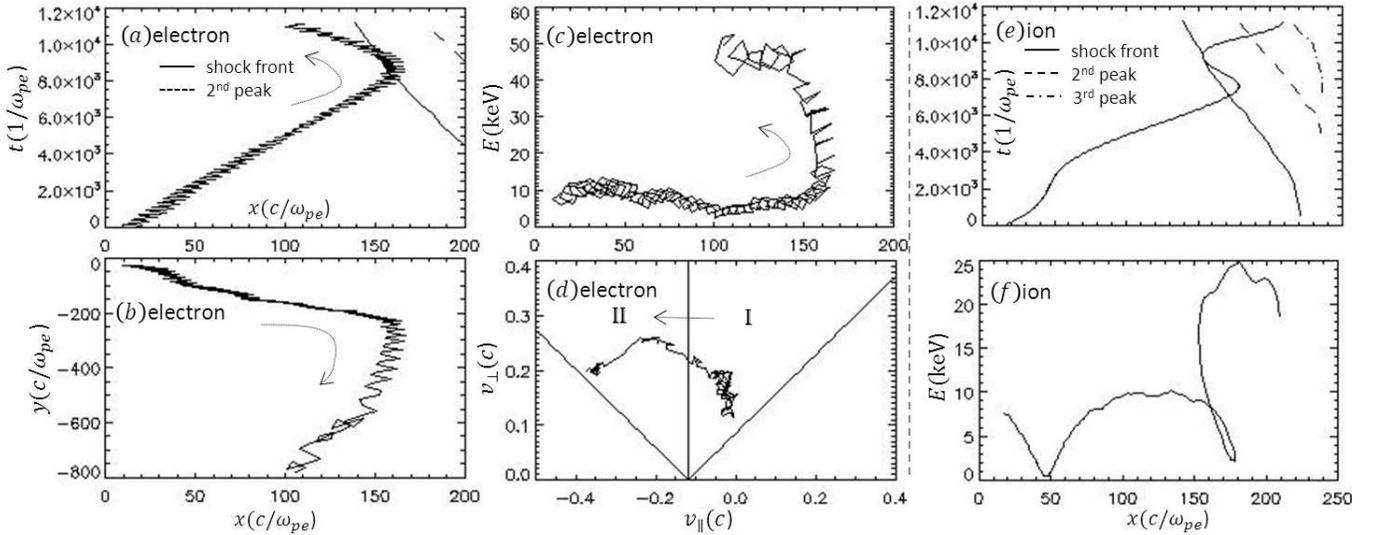}
\caption{(a-d): A typical electron tracking experiencing SDA.
(e\&f): A typical ion tracking experiencing SDA. }
\label{tracking}
\end{center}
\end{figure*}

In Figure \ref{edist}, we plot the electron and ion energy distributions, $f(E)(=dN/dE)$ upstream [$50<x<155 c/\omega_{pe}$ (b \& e)]  and downstream [$155<x<250 c/\omega_{pe}$ (c \& f)] at $t=8400/\omega_{pe}$ for $\theta_B=80^\circ$.
We fit the thermal (bulk) part of the distributions using a kappa distribution with $\kappa=10$ (dashed lines), yielding $T_e=T_i=0.8$keV  upstream and
$T_e=1.58$keV and $T_i=1.8$keV  downstream.
In Figure \ref{edist}(a\&d), we plot the electron and ion phase-space distributions in $E-x$, overlaid with $B_y$ field to indicate the location of the shock front. Abundant non-thermal electrons are seen ahead of the shock at $x=155 c/\omega_{pe}$, traveling as far back as $x=10c/\omega_{pe}$ [Figure \ref{edist}a].

In Figure \ref{edist}(b), the electron energy spectrum shows a deviation from a thermal distribution at $E\sim 12$keV and the spectrum becomes steeper at $E\sim 50$keV. The dotted line in Figure \ref{edist}(b) is the theoretical energy distribution via SDA when the electric potential energy jump $|e|\Delta\Phi\approx 3.5$(keV) at the shock front is considered (See Section \ref{eSDA}). In Figure \ref{edist}(e), the ion energy spectrum shows a deviation from a thermal distribution at $E\sim 15$keV and the spectrum becomes steeper at $E\sim 30$keV.

In Figure \ref{tracking}(a-d), we plot a typical track of an electron experiencing SDA, overlaid with the shock front and the subsequent compression peaks [Figure \ref{tracking}a]. 
The electron is reflected at the shock front [Figure \ref{tracking}a] and gains energy from $8$keV to $50$keV [Figure \ref{tracking}c]. After the reflection, the electron drifts along the upstream magnetic field lines [Figure \ref{tracking}b]. In Figure \ref{tracking}(d), we plot the electron in the $v_\perp-v_\parallel$ phase-space. The electrons in the region I are transferred to the region II after the reflection.

In Figure \ref{tracking}(e\&f), we plot a typical track for an ion experiencing SDA. Whether or not an ion gains energy via SDA depends on its incident speed and angle of incidence at the shock front \citep{kirk94}.
When the ion meets the shock front, it turns back toward the upstream
with a larger gyro-radius [Figure \ref{tracking}a] and is accelerated by the $E_z$ field. The kinetic energy of the ion increases from $5$keV up to $20$keV [Figure \ref{tracking}b]. A detailed analysis for ions experiencing SDA in  perpendicular shocks was described in our previous work \citep{park12}.

\subsubsection{Electron spectrum via SDA}
\label{eSDA}

\begin{figure}
\begin{center}
\includegraphics[scale=.34]{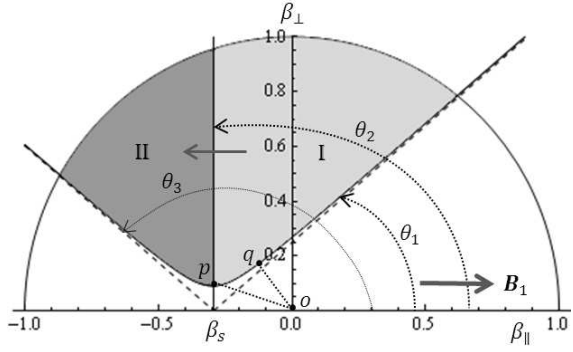}
\caption{The condition for SDA in the $\beta_\perp-\beta_\parallel$ space of the upstream rest frame when $\beta_s(=V_\text{sh}/c\text{cos}\theta_B)=-0.293$.
The incoming electrons in the region I (light shaded) are transferred to the region II (dark shaded) after reflection\citep{mann06,mann09}. The hyperbolic curve is given by the potential energy $e\Phi=-3.5$keV at the shock front, which approaches the straight dotted lines when $e\Phi \rightarrow 0$.}
\label{sda_condition}
\end{center}
\end{figure}

In this subsection, we generalize the electron energy spectrum via SDA by \citet{mann06,mann09,warmuth09} to include the electric potential energy,  $e\Phi$, at the shock front \citep[e.g.,][]{ball01}
and compare with the PIC simulation results.

First, we consider the de Hoffmann-Teller(dHT) frame (denoted by $'$) where the motional $E_z(=-V_1/cB_y)$ field vanishes.
The dHT frame is obtained by boosting with $v_s=V_{sh}/\text{cos}\theta_B$
along the magnetic field line in the upstream rest frame.
(Here we consider the negative shock speed, $V_{sh}<0$, for a shock traveling to the -$\hat{x}$ direction.)
The maximum $\theta_B$ for the existence of the dHT frame is given by $\theta_{B\text{max}}=\text{cos}^{-1}(V_{sh}/c)(<90^\circ)$.

In the dHT frame, the condition for an electron to reflect at the shock front is \citep[e.g.,][]{ball01,mann06,mann09}
\begin{equation}
\beta_{\perp}'>\beta_{\parallel}'\text{tan}\alpha_0 \,\,\,\text{and}\,\,\, \beta_{\parallel}'>0,
\label{sda_condi1}
\end{equation}
where $\perp$ and $\parallel$ are perpendicular and parallel to the upstream $\bold{B}_1$, respectively,
$\beta'_{\parallel,\perp}=v'_{\parallel,\perp}/c$, and $\alpha_0=\text{sin}^{-1}\sqrt{B_1/B_2}(\approx\text{sin}^{-1}\sqrt{1/r})$.
If we include the potential energy $e\Phi(<0)$ at the shock front, 
the reflection condition, Equation (\ref{sda_condi1}), can be written in the non-relativistic limit by \citep[e.g.,][]{ball01}
\begin{equation}
\beta_{\perp}'>\sqrt{\left(\beta'^2_{\parallel}-{2e\Phi\over m_e c^2}\right)}
\text{tan}\alpha_0, \,\,\, \beta_{\parallel}'>0.
\label{sda_condi2}
\end{equation}
For relativistic electrons with $\beta_{\parallel}'\gg \sqrt{2|e|\Phi/m_e c^2}(=0.117)$, the potential energy $e\Phi$ is negligible and Equation (\ref{sda_condi2}) is reduced to Equation (\ref{sda_condi1}).

Using the Lorentz transformation between the dHT frame($'$) and the upstream rest frame,
\begin{equation}
\beta'_\parallel={\beta_\parallel-\beta_s\over 1-\beta_\parallel \beta_s},
\,\,\,
\beta'_\perp={\beta_\perp\over \gamma_s(1-v_\parallel \beta_s)},
\label{lorentz1}
\end{equation}
where $\beta_s=v_s/c(=V_\text{sh}/c\text{cos}\theta_B)$ and $\gamma_s=1/\sqrt{1-\beta_s^2}$,
Equation (\ref{sda_condi2}) is transformed in the upstream rest frame into
\begin{align}
\beta_{\perp}&>\gamma_s\text{tan}\alpha_0
\left[(\beta_{\parallel}-\beta_s)^2-{2e\Phi\over m_ec^2}
(1-\beta_s\beta_\parallel)^2\right]^{1/2}, \nonumber \\
&\beta_\parallel>\beta_s.
\label{sda_condi3}
\end{align}

\begin{figure}
\begin{center}
\includegraphics[scale=.34]{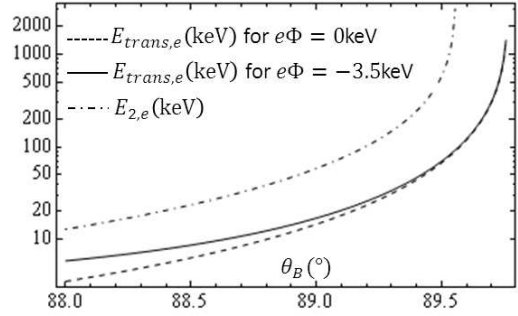}
\caption{The transition energy $E_{\text{trans},e}$ vs $\theta_B$ for $e\Phi=-3.5$ keV (solid) and $0$ keV (dashed) from Eq.(\ref{etrans}).
The energy point $E_{2,e}$ vs $\theta_B$ (dot-dashed) from Eq.(\ref{e2}).
Here $V_{sh}=0.0041c$, $r=2.5$ and $m_i/m_e=1836$ ($M_A=6.62$ and $\beta_p=8.93$.)}
\label{etrans_fig}
\end{center}
\end{figure}

The velocity of the electron after reflection in the dHT frame is given by
$\beta_{r\parallel}'=-\beta_{i\parallel}'$ and 
$\beta_{r\perp}'=\beta_{i\perp}'$, where the indices $i$ and $r$ represent the incoming and reflected electron, respectively. In the upstream rest frame, one gets \citep{mann06,mann09}
\begin{align}
\beta_{i\parallel}={2\beta_s-\beta_{r\parallel} (1+\beta_s^2)
\over 1-2\beta_{r\parallel}\beta_s+\beta_s^2}, 
\beta_{i\perp}={\beta_{r\perp}\over \gamma_s^2\left(1-2\beta_{r\parallel} \beta_s+\beta_s^2\right)}.
\label{vel_re}
\end{align}

\begin{figure*}
\begin{center}
\includegraphics[scale=.36]{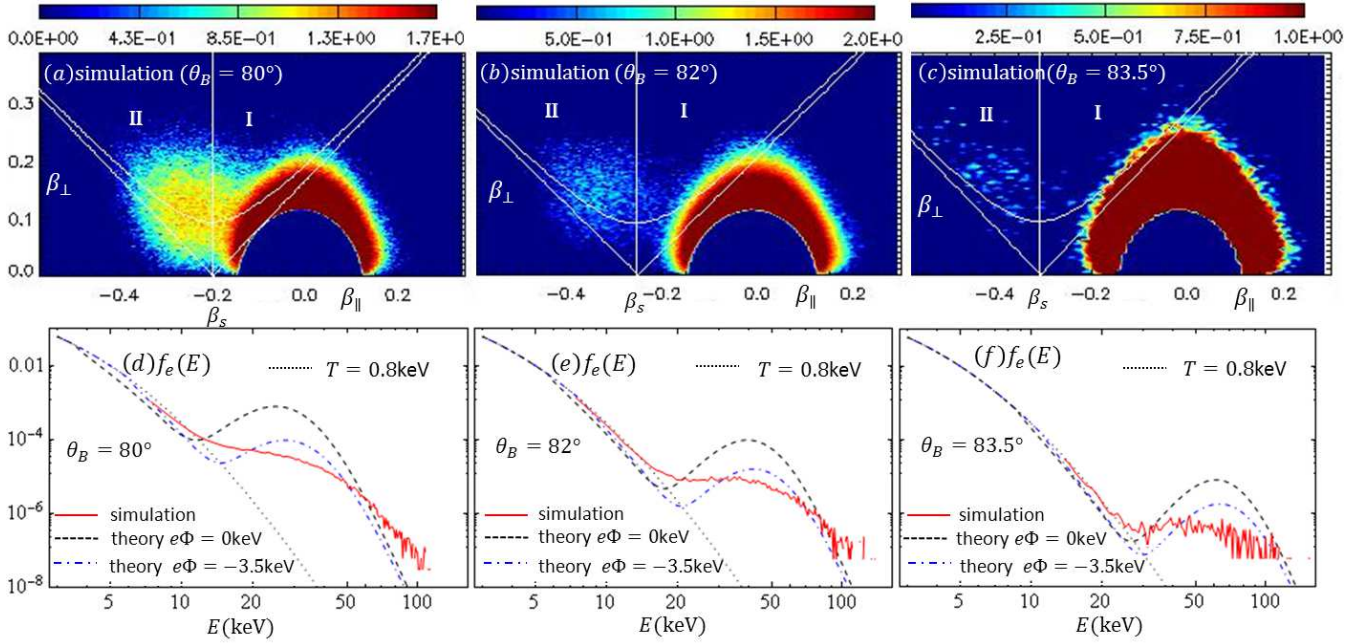}
\caption{(a-c): Simulation results of the upstream electron distribution in the $\beta_\perp-\beta_\parallel$ space of the upstream rest frame for $\theta_B=80^\circ, 82^\circ,$ and $83.5^\circ$.
(d$\sim$f): The normalized upstream electron energy distributions in the upstream rest frame from the simulation (solid), and the theory with $e\Phi=0$ (dashed) and $e\Phi=-3.5$keV (dot-dashed) at the shock front. 
We fit the thermal distribution with a kappa distribution with $T=0.8$keV and $\kappa=10$ (dotted).
Here $V_{sh}=-0.032c$, $r=2.5$ and $m_i/m_e=30$ ($M_A=6.62$ and $\beta_p=8.93$).}
\label{sda}
\end{center}
\end{figure*}

In Figure \ref{sda_condition}, we plot the SDA condition in the $\beta_\perp-\beta_\parallel$ space for $\beta_s=-0.293$ as an example \citep[e.g.,][]{mann06,mann09}.
Equation (\ref{sda_condi3}) defines the region I (light shaded region) and the hyperbolic curve is given by the potential energy $e\Phi=-3.5$keV at the shock front. When $e\Phi$ goes to zero, the hyperbolic curve approaches the straight dotted lines as seen in \citet{mann06,mann09}.
Equation (\ref{vel_re}) implies that the incoming electrons in the region I are transferred to the region II (dark shaded region) after reflection.
As $\theta_B$ increases to $\theta_{B\text{max}}(=\text{cos}^{-1}(V_{sh}/c))$, $\beta_s$ goes to $-1$ in Figure \ref{sda_condition} and the number of electrons satisfying the reflection condition in Equation (\ref{sda_condi3}) decrease to zero. Therefore, no electron reflects at the shock front in the superluminal shocks (where $|\beta_s|\ge 1$) or the perpendicular shocks.

The threshold energy $E_\text{thres}$ for SDA is given by the shortest distance
from the origin to the hyperbolic curve, $\overline{oq}(=\beta_{\text{thres}})$ in Figure \ref{sda_condition}, namely
\begin{align}
E_\text{thres}&=\left(1/\sqrt{1-\beta_\text{thres}^2}-1\right)m_ec^2, 
\label{thres} \,\, \text{where} \\
\beta_{\text{thres}}&=\sqrt{{D^2\left[(1-\beta_s)^2D^2P^2
\{P-(1+\beta_s^2)\}+\beta_s^2-P\right]\over 1+D^2(1-P\beta_s^2)}}, \nonumber
\end{align}
where $D=\gamma_s\text{tan}\alpha_0$ and $P=2e\Phi/m_ec^2$.

The transition energy $E_{\text{trans},e}$ between thermal and non-thermal electron populations 
is given by the distance from the origin to the point $p$, $\overline{op}(=\beta_{\text{trans}})$, in Figure \ref{sda_condition} such as
\begin{align}
E_{\text{trans},e}&=\left(1/\sqrt{1-\beta_\text{trans}^2}-1\right)m_ec^2, 
\label{etrans} \,\,\text{where} \\
\beta_{\text{trans}}&=\beta_s^2-(1-\beta_s^2)(2e\Phi/m_ec^2)\text{tan}^2\alpha_0. \nonumber
\end{align}

The energy point $E_{2,e}$ where the maximum energy ratio of the reflected to the incoming electron occurs is given by \citet{ball01} in the non-relativistic approximation, namely
\begin{align}
E_{2,e}&\approx\left(1/\sqrt{1-\beta_2^2}-1\right)m_ec^2, \nonumber \\
\beta_2&=2\beta_s\text{cos}(\alpha_0/2),
\label{e2}
\end{align}
Beyond $E=E_{2,e}$, the spectral index $\delta$, defined as in $f(E)\propto E^{-\delta}$, increases. 
Figure \ref{etrans_fig} shows how the transition energy $E_{\text{trans},e}$ for $e\Phi=-3.5$ keV (solid) and $0$ keV (dashed), and $E_{2,e}$ (dot-dashed) varies with $\theta_B$ when $V_{sh}=0.0041c$, $r=2.5$ and $m_i/m_e=1836$ ($M_A=6.62$ and $\beta_p=8.93$).

The reflected electron distribution in the upstream rest frame is written as
\begin{align}
f_r(\boldsymbol{\beta}_r)=f_i[\boldsymbol{\beta}_i(\boldsymbol{\beta}_r)](d^3\beta_i/d^3\beta_r)
\Theta(\beta_s-\beta_{r\parallel})\times \nonumber \\
\Theta\left(\beta_{r\perp}-\gamma_s\text{tan}\alpha_0
\left[(\beta_s-\beta_{r\parallel})^2-{2e\Phi\over m_ec^2}
(1-\beta_s\beta_{r\parallel})^2\right]^{1/2}\right),
\label{fdist_sda}
\end{align}
where $\Theta$ is the step function and the term ${d^3\beta_i/d^3\beta_r}$ is given by the Jacobian determinant,
\begin{equation}
{d^3\beta_i\over d^3\beta_r}={\beta_{i\perp}\over\beta_{r\perp}}
\left|{\partial(\beta_{i\perp},\beta_{i\parallel})\over \partial(\beta_{r\perp},\beta_{r\parallel})}\right|=
{(1-\beta_s)^4\over (1-2\beta_{r\parallel} \beta_s+\beta_s^2)^{4}}.
\label{jacobi}
\end{equation}
Here we let the incoming distribution in Equation (\ref{fdist_sda}) be a semi-relativistic kappa distribution \citep{mann06,mann09}, 
\begin{equation}
f_i(\boldsymbol{\beta}_i)=c_\kappa\left(1+{2(\gamma_i-1)\over 2\kappa-3}{m_ec^2\over T}\right)^{-\kappa-1},
\label{fidist}
\end{equation}
where $\gamma_i=1/\sqrt{1-\boldsymbol{\beta}_i^2}$ and $c_\kappa=1/\int d^3\beta_i f_i(\boldsymbol{\beta}_i)$.

We calculate the upstream electron energy distributions,
$f(E)=f(\beta)d\beta/dE$, in the upstream rest frame using Equations ({\ref{vel_re}-\ref{fidist}) to obtain
\begin{align}
f(\beta)=\left\{\begin{array}{cc}
2\pi\beta^2\int_{-1}^{1}dt f_i({\beta}) \\
\mbox{.................   for}\,\,\beta<\beta_\text{thres} \\
2\pi\beta^2\left(\int_{-1}^{t_3}dt+\int_{t_1}^{1}dt\right)f_i(\beta) \\
\mbox{.................    for}\,\,\beta_\text{thres}\leq\beta<\beta_\text{trans}, \\
2\pi\beta^2\int_{t_3}^{t_2}dt {d^3\beta_i\over d^3\beta_r} 
f_i[\boldsymbol{\beta}_i(\beta,t)] \\
+2\pi\beta^2\left(\int_{-1}^{t_2}dt+\int_{t_1}^{1}dt\right)f_i(\beta) \\
\mbox{.................    for}\,\,\beta\geq\beta_\text{trans}
\end{array}\right.
\label{fe_theory}
\end{align}
where $t=\text{cos}\theta$ and $\theta$ is the electron's pitch angle, i.e., the angle between $\boldsymbol{\beta}$ and $\bold{B}_1$. 
The boundaries $t_1(=\text{cos}\theta_1)$, $t_2(=\text{cos}\theta_2)$, and
$t_3(=\text{cos}\theta_3)$ as shown in Figure \ref{sda_condition} are given as $t_2=\beta_s/\beta$, and roots of the equation,
\begin{align}
t_{1,3}^2\left\{1+D^2\left(1-P\beta_s^2\right)\right\}
-2t_{1,3} D^2{\beta_s\over\beta}\left(1-P\right) \nonumber \\
+{D^2\over\beta^2}
\left(\beta_s^2-P\right)-1 =0.
\end{align}

\begin{figure}
\begin{center}
\includegraphics[scale=.38]{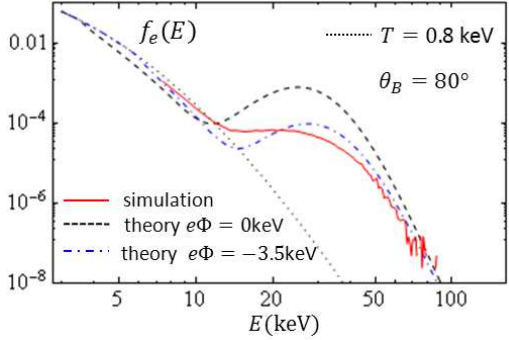}
\caption{The normalized upstream electron energy distribution in the upstream rest frame from the out-of-plane upstream $B$-field simulation (solid) ($\bold{B}_1=B_{1x}\hat{x}+B_{1z}\hat{z}$). 
The dashed and dot-dashed lines are theoretical results for
$e\Phi=0$ and $e\Phi=-3.5$keV at the shock front, respectively.
}
\label{fdist-out}
\end{center}
\end{figure}

In Figure \ref{sda}(a-c), we plot the upstream electron distribution in the $\beta_\perp-\beta_\parallel$ space of the upstream rest frame from the simulations for $\theta_B=80,82$ and $83.5^\circ$.
Electrons in  region I are transferred to  region II. 
As $\theta_B$ increases, the number of electrons participating in SDA 
decreases since the energy threshold for SDA in Equation (\ref{thres}) increases.

In Figure \ref{sda}(d$\sim$f), we plot the normalized upstream electron energy distributions, $f(E)(=1/NdN/dE)$, in the upstream rest frame for $\theta_B=80,82$ and $83.5^\circ$.
We compare the results of Equation (\ref{fe_theory}) for $e\Phi=0$ (dashed) and $-3.5$(keV) (dot-dashed) with the simulation result (solid).
The dotted line is the incoming kappa distribution with $T=0.8$keV and $\kappa=10$.
Using Equation (\ref{etrans}), the transition energy points are given by 
$E_{\text{trans},e}=11.3$keV, $16.5$keV, and $24.2$keV for the angles, $\theta_B=80^\circ$, $82^\circ$, and $83.5^\circ$, respectively, when $e\Phi=-3.5$keV. 
Using Equation (\ref{e2}), the energy points $E_{2,e}$'s, beyond which the spectral index increases, are given by
$E_{2,e}=34$keV, $56$keV, and $93$keV for the angles, $\theta_B=80^\circ$, $82^\circ$, and $83.5^\circ$, respectively.
Here $V_{sh}=-0.032c$, $r=2.5$ and $m_i/m_e=30$ ($M_A=6.62$ and $\beta_p=8.93$).

In Figure \ref{fdist-out}, we plot the upstream electron energy distribution in the upstream rest frame from the out-of-plane upstream $B$-field simulation (solid) where $\bold{B}_1=B_{1x}\hat{x}+B_{1z}\hat{z}$ and $\theta_B=80^\circ$.
The dashed and dot-dashed lines are the theoretical results for
$e\Phi=0$ and $e\Phi=-3.5$keV at the shock front, respectively. 
In this simulation, the temperature anisotropy-driven instabilities seen in Figure \ref{fft} are precluded. By direct comparison with the electron energy distribution in Figure \ref{fdist_sda}(d), the simulation result in Figure \ref{fdist-out} is closer to the theoretical result for $e\Phi=-3.5$keV. 
That is, the high energy tail is steeper in Figure \ref{fdist-out}. 
This implies that the temperature anisotropy-driven instabilities near the shock front can contribute to further electron acceleration, possibly via the interaction between electrons and perturbed magnetic fields.
There is a still discrepancy between the theory with $e\Phi=-3.5$keV and the simulation around the transition energy, $E_{\text{trans},e}=11.3$keV. This is probably due to the effects of small-scale waves generated in the transition region as indicated by \citet{matsukiyo11}.

\subsubsection{Bremsstrahlung radiation}

\begin{figure*}
\begin{center}
\includegraphics[scale=.4]{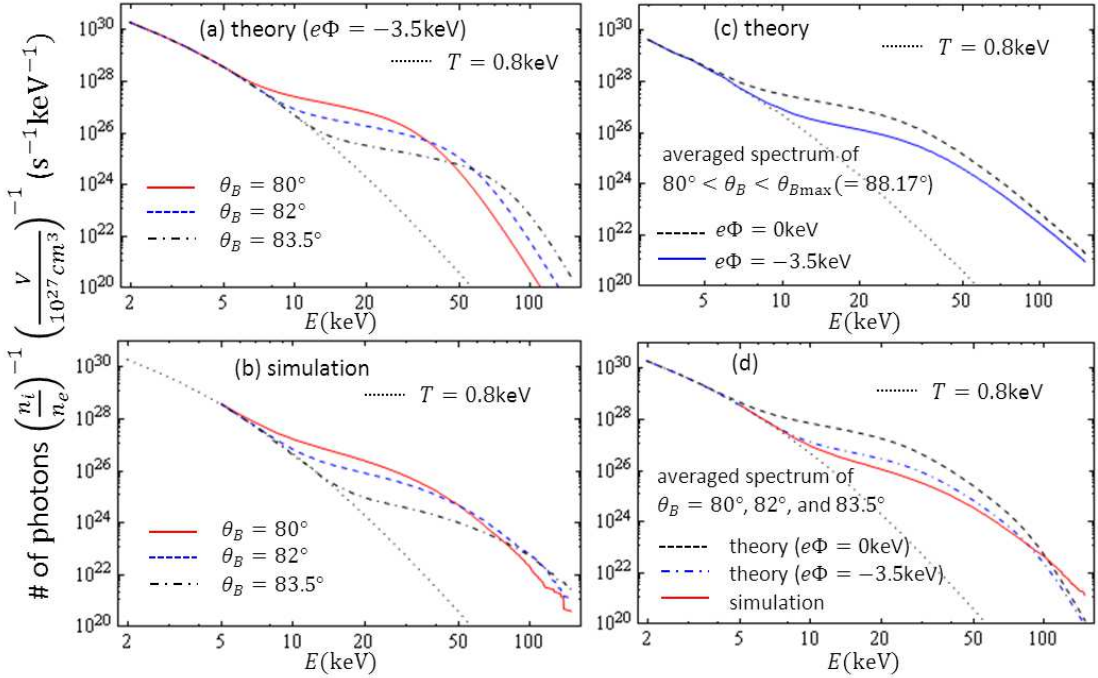}
\caption{(a\&b): The photon distribution via bremsstrahlung radiation
for $\theta_B=80^\circ$, $82^\circ$, and $83.5^\circ$ 
when the electron distributions are given by Eq.(\ref{fe_theory}) and by simulations. 
(c): The averaged photon distribution of $80^\circ\le\theta_B\le\theta_{B\text{max}}(=88.17^\circ)$ from the theoretical distributions with $e\Phi=0$ and $-3.5$keV.
(d): The averaged photon distribution of $\theta_B=80^\circ$, $82^\circ$, and $83.5^\circ$. The dotted line is the photon distribution when the electron distribution is a kappa distribution with $T=0.8$keV and $\kappa=10$.
Here $V_{sh}=-0.032c$, $r=2.5$ and $m_i/m_e=30$ ($M_A=6.62$ and $\beta_p=8.93$).}
\label{bremsst}
\end{center}
\end{figure*}

Electrons accelerated via SDA move along the magnetic field lines as can be  seen from  the electron tracking in Figure \ref{tracking}(b) and will collide with ions to emit bremsstrahlung radiation.  
In the low-frequency limit of $\omega b/\gamma v\ll 1$, where $\omega$ is a photon angular frequency, $b$ is the impact parameter, $v$ is the electron speed, and $\gamma=1/\sqrt{1-(v/c)^2}$,
the number of photons per unit frequency per unit volume per unit time produced by the bremsstrahlung process is \citep[e.g.,][]{rybicki79}
\begin{equation}
{dN\over d\omega dV dt}={16 Z^2e^6n_en_i\over 3c^3m_e^2\hbar^2\omega}
\int_{v_\text{min}}^{c} {\gamma\over v}\text{ln}\left({\gamma m_ev^2\over\hbar\omega}\right)f(v)dv,
\label{flux}
\end{equation}
where $f(v)$ is a normalized electron distribution and $v_\text{min}$ is 
determined by the equation $\hbar\omega=(\gamma_\text{min}-1)mc^2$.

Given the electron distributions in Figure \ref{sda}(d-f),
we calculate the number of photons per unit time(sec) per unit energy(keV) using Equation (\ref{flux}) for the ion density of $n_i=n_e$ and the volume of the region emitting the X-rays, $V=10^{27}\text{cm}^3$.
In Figure \ref{bremsst}(a\&b), we show the photon distribution for $\theta_B=80$ (solid), $82$ (dashed), and $83.5$ (dot-dashed) when the electron distribution is given by the theoretical distribution in Equation (\ref{fe_theory}) with $e\Phi=-3.5$keV [Figure \ref{bremsst}a] and by the simulation [Figure \ref{bremsst}b].
The dotted line is the photon distribution when the electron distribution is given by a kappa distribution with $T=0.8$keV and $\kappa=10$.
Here $V_{sh}=-0.032c$, $r=2.5$ and $m_i/m_e=30$ ($M_A=6.62$ and $\beta_p=8.93$).

In Figure \ref{bremsst}(c), we plot the averaged photon distribution of  $80^\circ\le\theta_B\le\theta_{B\text{max}}(=88.17^\circ)$ 
from the theoretical distribution in Equation (\ref{fe_theory}) with $e\Phi=0$ (dashed) and $e\Phi=-3.5$keV (solid).
In Figure \ref{bremsst}(d), we plot the averaged photon distribution of $\theta_B=80,$ $82$, and $83.5^\circ$ from the electron distribution given by Equation (\ref{fe_theory}) with $e\Phi=0$ (dashed) and $e\Phi=-3.5$keV (dot-dashed), and from the simulations (solid).
In Figure \ref{bremsst}(d), we notice that the theoretical result with $e\Phi=-3.5$keV (dot-dashed) is in good agreement with the simulation result (solid).

In Figure \ref{bremsst}(d), the simulation result shows that
the transition energy between the thermal and non-thermal photon spectrum is $E_{\text{trans},p}\approx 10$keV and
the energy point beyond which the photon spectrum becomes steeper, is $E_{2,p}\approx 40$keV. 
The spectral index is $\delta=3$(simulation) and $\delta=2.2$(theory with $e\Phi=-3.5$keV) in $10$keV$<E<40$keV  
and $\delta=7$(simulation) in $E>40$keV.
For emission from multiple shocks with different $\theta_B$'s, 
the transition energy $E_{\text{trans},p}$ and $E_{2,p}$ 
would be dominated by the shock with the minimum $\theta_B$.
Note that the transition energy $E_{\text{trans},p}$ for the photon distribution is a bit smaller than $E_{\text{trans},e}$ for the electron distribution because bremsstrahlung photons are produced by electrons with higher energies than the photon energy \citep[e.g.,][]{holman03}.
The energy point $E_{2,p}$ is approximately given by Equation (\ref{e2}).

RHESSI data for several solar flares (Table I in \citet{altyntsev12}), shows that the spectral index $\delta$ is in the range $2.5<\delta<3$ and the transition energy is in the range 12.1 keV $<E_{\text{trans},p}<29.2$ keV. 
Therefore, the electron energization via SDA well explains some of the RHESSI X-ray spectra for the energy regime, $E<E_{2,p}$
and how the transition energy is related with the shock geometry, i.e., the minimum $\theta_B$.

The observed RHESSI spectra do not show a steepening beyond $E=E_{2,p}$, and thus the theory herein cannot by itself account for the electron acceleration to produce those photons.
This indicates that additional mechanisms, such as diffusive shock acceleration,  are required for further electron energization to maintain the power-law spectrum up to $E\sim$ MeV. 
It is not unreasonable to expect that solar flares involve multiple acceleration mechanisms operating on a range of scales.

\subsection{Effects of a realistic proton/electron mass ratio on the spectra} 

\begin{figure}
\begin{center}
\includegraphics[scale=.38]{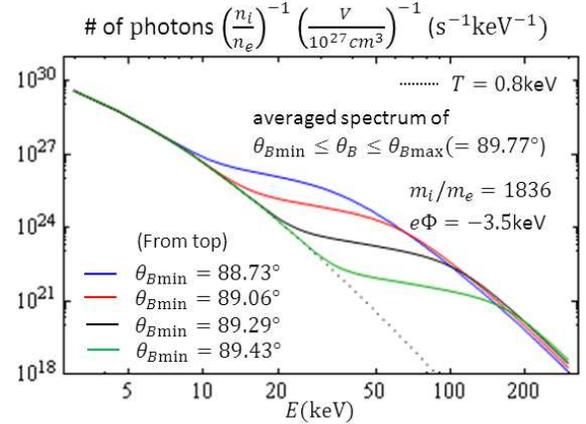}
\caption{The averaged photon distribution of $\theta_{B\text{min}}\le\theta_B\le\theta_{B\text{max}}(=89.77^\circ)$ from the theoretical distributions in Equation (\ref{fe_theory}) with $e\Phi=-3.5$keV for several minimum values of $\theta_B$. Here $V_{sh}=0.0041c$, $r=2.5$ and $m_i/m_e=1836$ ($M_A=6.62$ and $\beta_p=8.93$). 
The transition energy increases as $\theta_{B\text{min}}$ increases.}
\label{realmass}
\end{center}
\end{figure}

With the actual ion/electron mass ratio of $m_i/m_e=1836$,
the shock speed in the upstream rest frame is reduced by $\sqrt{30/1836}$ compared to our simulations when the Mach number \textit{M} and the plasma $\beta_p$ are fixed.
The compression ratio $r$ and the electric potential energy $e\Phi$ are unchanged for a fixed \textit{M} and $\beta_p$ \citep[e.g.,][]{hoshino01}.
Therefore, the shock structure are not expected to change for a realistic mass ratio simulation.

The electron energy spectrum in Equation (\ref{vel_re}) depends only on 
$\beta_s$, $\alpha_0$, and $e\Phi$.
For a fixed $M$ and $\beta_p$, $\alpha_0(\equiv \text{sin}^{-1}\sqrt{B_1/B_2})$ and $e\Phi$ are unchanged.
From the definition of $\beta_s(\equiv V_{sh}/c\text{cos}\theta_B)$,
we only need to change the angle from $\theta_B$ into $\theta'_B$ 
to obtain the electron energy spectrum for the mass ratio $m_i/m_e=1836$ from that with $m_i/m_e=30$, 
\begin{equation}
\theta'_B=\text{cos}^{-1}\left[{\text{cos}\theta_B\over\sqrt{1836/30}}\right].
\end{equation}
For example, the averaged spectrum with $80^\circ\le\theta_B\le\theta_{B\text{max}}(=88.17^\circ)$ for $m_i/m_e=30$ corresponds to that with $88.73^\circ\le\theta_B\le 89.77^\circ$ for $m_i/m_e=1836$.

In Figure \ref{realmass}, we plot the averaged photon distribution of $\theta_{B\text{min}}\le\theta_B\le\theta_{B\text{max}}(=89.77^\circ)$ for a real ion/electron mass ratio, $m_i/m_e=1836$, 
from the electron distribution given by Equation (\ref{fe_theory}) with $e\Phi=-3.5$keV.  
Here $V_{sh}=0.0041c$ and $r=2.5$ ($M_A=6.62$ and $\beta_p=8.93$). 
The different values of $\theta_{B\text{min}}$, from 
$\theta_{B\text{min}}=88.73^\circ$ to $89.43^\circ$, give the different transition energy points, from $E_{\text{trans},p}=10$ to $35$ keV. The power indices of the photon spectrum are nearly the same as $\delta\sim 2.2$ in $E_{\text{trans},p}<E<E_{2,p}$, where $E_{2,p}$ runs from $40$ to $150$ keV.

\section{conclusion}
\label{summary}

In summary, we studied quasi-perpendicular, low $M$/high $\beta_p$ shocks with full PIC 2D simulations using a reduced ion/electron mass ratio $m_i/m_e=30$. The shock compression ratio we found was in agreement with the Rankine-Hugoniot relation. Whistler instabilities driven by downstream temperature anisotropy were observed. A modified two-stream instability due to the incoming and reflecting ions in the shock transition region was also observed.

Abundant non-thermal electrons accelerated via SDA were observed upstream. 
We compared the electron energy distribution from the simulations with the distributions derived by extending a theoretical model \citep{mann06,mann09,warmuth09} and found that they reasonably agree with each other. 

In the perpendicular shocks, however, SDA can be achieved only by particles transmitting into the downstream, and their energy gains are smaller than those of the reflected ones in quasi-perpendicular shocks here \citep[e.g.,][]{ball01}.
Therefore, such abundant non-thermal electrons observed in this paper were not seen in the perpendicular shocks \cite[e.g.,][]{park12}.

We calculated the photon flux via bremsstrahlung radiation from the electron distributions from both the theory and the simulations.  We showed that a transition energy, $E_{\text{trans},p}$, marking the transition  from a thermal to a non-thermal part of the photon spectrum, is determined by the minimum $\theta_B$ from multiple shocks with different $\theta_B$'s in $\theta_{B\text{min}}\le\theta_B\le 90^\circ$. 
Different solar flares have different $\theta_{B\text{min}}$'s in their termination shocks and therefore can show different transition energy points.

From the simulations, the averaged photon spectrum of $\theta_B=80^\circ,$ $82^\circ$, and $83.5^\circ$ gives a spectral index $\delta\sim 3$ in $10$keV$<E<40$keV and the spectral index increases beyond $E=40$keV. 
The spectral index $\delta\sim 3$ as well as the transition energy, $E_{\text{trans},p}=10$keV, well explains some of the RHESSI X-ray spectra in the energy regime, $E<40$keV. To account for the spectral index of the RHESSI X-ray spectrum up to $E\sim$MeV, however, additional mechanisms other than SDA are required.

Note that although our simulations were performed using  $m_i/m_e=30$, we  analytically scaled the results to the realistic ion/electron mass ratio, and  found that the predicted photon spectra are  indeed insensitive to this mass ratio.

\begin{acknowledgments}
This work was supported by NSF under Grant PHY-0903797, by DOE under Grant No. DE-FG02-06ER54879 and Cooperate Agreement No. DE-FC52-08NA28302, and by NSFC under Grant No. 11129503.
We also thank the OSIRIS consortium for the use of OSIRIS.
The research used resources of NERSC. 
\end{acknowledgments}

\appendix

\section{generation of a kappa distribution}
\label{app1}

To generate a kappa distribution, we use the random number distribution by \citet{leitner11}.
We denote $\{N_i(0,1)|i=1,2,...,N\}$ as a set of normally distributed random numbers with the mean of $0$ and the deviation of $1$, 
and $\{U_i(0,1)|i=1,2,...,N\}$ as a set of uniformly distributed random numbers between [0,1].
Then a sequence of the random number $x_i$ given by
\begin{equation}
x_i=b_1 N_i(0,1)+b_2U_i(0,1)x_{i-1}
\label{random}
\end{equation}
generates a 1D kappa distribution with the mean of $0$ and the deviation of $1$, and the coefficients, $b_1$ and $b_2$ determine the index $\kappa$ \citep [Figure 8 in][]{leitner11}. 

To implement the kappa generator in OSIRIS, we determine 
the particle's initial momentum $\bold{p}$ as 
\begin{equation}
\bold{p}_i=\bold{p}_{th} x_i+\bold{p}_{d},
\label{prandom}
\end{equation}
where $x_i$ is given by Equation (\ref{random}), $\bold{p}_{th}$ and $\bold{p}_{d}$ are the thermal and the drift momentum, respectively. For $\kappa=10$, we choose $b_1=0.58$ and $b_2=1.15$. Then a sequence of $\bold{p}_i$ in Equation (\ref{prandom}) generates a 3D kappa distribution in Figure \ref{kappadist}.

\section{dispersion relation for the modified two-stream instability}
\label{app2}

The kinetic dispersion relation for electrostatic instabilities is
\begin{equation}
1+\sum_{s=e,i}K_s(\bold{k},\omega)=0,
\end{equation}
where $K_s(\bold{k},\omega)$ is the susceptibility.
For magnetized electrons with isotropic Maxwellian distributions,
$K_e(\bold{k},\omega)$ is given by \citep{gary93}
\begin{equation}
K_e(\bold{k},\omega)={\omega_{pe}^2\over v_{eth}^2 k^2}
\left[1+\xi_e^0 e^{-\lambda_e}
\sum_{m=-\infty}^{\infty}I_m(\lambda_e)Z(\xi_e^m)\right],
\label{suscept1}
\end{equation}
where $v_{eth}$ is the thermal velocity, $I_m$ is the modified Bessel function of the first kind, $\lambda_e=k_\perp^2v_{eth}^2/\Omega_e^2$, $\Omega_e=eB/(m_ec)(<0)$, $Z$ is the plasma dispersion function, and $\xi_e^m=(\omega-m\Omega_e)/(\sqrt{2}k_\parallel v_{eth})$.
For unmagnetized ions with drifting Maxwellian distributions,
\begin{equation}
K_i(\bold{k},\omega)=-{\omega_{pi}^2\over 2 v_{ith}^2 k^2}Z'(\xi_i),
\end{equation}
where $\xi_i=(\omega-\bold{k}\cdot\bold{V}_{id})/(\sqrt{2}k v_{ith})$
and $'$ is the derivative with respect to $\xi_s$.

Here we consider that $\bold{B}\approx B_y\hat{y}$, $\bold{k}=k_x\hat{x}$
$\bold{V}_{sd}=V_{sd}\hat{x}$.
Then the term $\xi_e^0 Z(\xi_s^m)$ in Equation (\ref{suscept1}) becomes $-\omega/(\omega-m\Omega_e)$ as $k_\parallel$ goes to $0$.
The dispersion relation for the MTSI in the electrostatic limit becomes
\begin{equation}
1+{\omega_{pe}^2\over v_{eth}^2 k^2}
\left(1-e^{-\lambda_e}\sum_{m=-\infty}^{\infty}I_m(\lambda_e)
{\omega\over\omega-m\Omega_e}\right)
-\sum_{s=in,re}{\omega_{ps}^2\over 2 v_{sth}^2 k^2}Z'(\xi_s)=0.
\label{mtsi}
\end{equation}

\singlespace
\bibliographystyle{apj}

\end{document}